\documentclass{aastex}
\usepackage{spr-astr-addons}
\usepackage{url}

\RequirePackage{color}

\usepackage{latexsym}
\usepackage{graphics}
\usepackage{graphicx}
\usepackage{epsfig}

\begin{document}
\title{rp-process weak-interaction mediated rates of waiting-point nuclei}
\shorttitle{rp-process weak rates}
\author{Jameel-Un Nabi\altaffilmark{1}}
\affil{Faculty of Engineering Sciences, GIK Institute of Engineering
Sciences and Technology, Topi 23640, Swabi, Khyber Pakhtunkhwa,
Pakistan\\ email: jameel@giki.edu.pk }
\altaffiltext{1}{The Abdus
Salam ICTP, Strada Costiera 11, 34014, Trieste, Italy}
\begin{abstract}
Electron capture and positron decay rates are calculated for
neutron-deficient Kr and Sr waiting point nuclei in stellar matter.
The calculation is performed within the framework of pn-QRPA model
for rp-process conditions. Fine tuning of particle-particle,
particle-hole interaction parameters and a proper choice of the
deformation parameter resulted in an accurate reproduction of the
measured half-lives. The same model parameters were used to
calculate stellar rates. Inclusion of measured Gamow-Teller strength
distributions finally led  to a reliable calculation of weak rates
that reproduced the measured half-lives well under limiting
conditions. For the rp-process conditions, electron capture and
positron decay rates on $^{72}$Kr and $^{76}$Sr are of comparable
magnitude whereas electron capture rates on $^{78}$Sr and $^{74}$Kr
are 1--2 orders of magnitude bigger than the corresponding positron
decay rates. The pn-QRPA calculated electron capture rates on
$^{74}$Kr are bigger than previously calculated. The present
calculation strongly suggests that, under rp-process conditions,
electron capture rates form an integral part of weak-interaction
mediated rates and should not be neglected in nuclear reaction
network calculations as done previously.
\end{abstract}
\section{Introduction}
\label{intro} At low temperatures (T$_{9} \leq 0.3 $K), energy
generation and accompanying nucleosynthesis in explosive hydrogen
burning are characterized by the hot CNO cycles and at higher
temperatures by the rp- and the $\alpha$p-processes (see for example
\cite{Pri00}). Recent studies (e.g. \cite{Wor94}) have pointed
toward extreme hydrogen burning where at sufficiently high
temperature (T$_{9} \leq 0.8 $K) and density conditions ($\rho \geq
10^{4}$ gcm$^{-3}$), depending on the time scale of the explosive
event, the rp- and the $\alpha$p-processes reaction path may well
proceed beyond mass A =64 and Z = 32.

The rp-process is characterized by proton capture reaction rates
that are orders of magnitude faster than any other competing
process, specially $\beta$-decay rates. The reaction path follows a
series of fast (p,$\gamma$)-reactions until further proton capture
is inhibited, either by negative proton capture Q-values (proton
decay) or small positive proton capture Q-values
(photodisintegration). Now the reaction flow has to wait for the
relatively slow $\beta$-decay and the respective nucleus is hence
called a waiting point. It is primarily the reliable estimate of
half-lives of the waiting point nuclei that ultimately determine the
timescale of the nucleosynthesis process and the produced isotopic
abundances. The rp-process produces rapid nucleosynthesis on the
proton-rich side of stability line toward heavier proton-rich
nuclei. Such processes typically occur when hydrogen fuel is ignited
under highly degenerate conditions in explosive events on the
surface of compact objects like white dwarfs (novae) and neutron
stars (X-ray bursts).

Type I X-ray bursts have been pointed out as possible sites for high
temperature hydrogen burning via the rp-process \cite{Sch98}. The
X-ray bursts are typically generated by a thermonuclear runaway in
the hydrogen-rich environment of an accreting compact object fed
from a binary companion. The timescale for the thermal runaway
ranges from 10--100 s \cite{Wor94}. Within this timescale, the
rp-process may proceed well beyond $^{56}$Ni \cite{Wor94}. Previous
attempts to extend the network calculations beyond Z = 32 suffered
considerably due to lack of reliable nuclear physics data
\cite{Sch98}. It was pointed out by authors in Ref. \cite{Sch98}
that important parameters for a successful rp-process
nucleosynthesis calculation include nuclear masses, nuclear
deformations (specially in the regime A = 70--80 because of the wide
variety of nuclear shapes displayed in the region) and finally the
reliable calculation of stellar electron capture and $\beta$-decay
rates of the waiting point nuclei along a given reaction path since
they determine time structure and abundance patterns.

This article reports on the microscopic calculation of
weak-interaction mediated rates (electron capture and $\beta$-decay
rates) on waiting point nuclei $^{72,74}$Kr and $^{76,78}$Sr. (It is
to be noted that in a true sense only the even-even $N$ = $Z$
nuclei, $^{72}$Kr and $^{76}$Sr, act as waiting point nuclei because
the next nucleus in the proton-capture chain is proton unbound.
However, at times, the deformed isotopes, $^{74}$Kr and $^{78}$Sr,
are also considered as waiting point nuclei in literature (see e.g.
\cite{Sar09}). It is in this sense that  $^{72,74}$Kr and
$^{76,78}$Sr are termed as waiting point nuclei in this manuscript.)
The nuclear model chosen to perform this task is the proton-neutron
quasiparticle random phase approximation (pn-QRPA) which has a
proven track-record for the calculation of electron capture and
$\beta$-decay rates. Half-lives of $\beta^{-}$ decays were
calculated systematically for about 6000 neutron-rich nuclei between
the beta stability line and the neutron drip line using the pn-QRPA
model \cite{Sta90}. Similarly half-lives for $\beta^{+}$/EC
(electron capture) decays for neutron-deficient nuclei with atomic
numbers Z = 10 - 108 were calculated up to the proton drip line for
more than 2000 nuclei using the same model \cite{Hir93}. These
microscopic calculations gave a remarkably good agreement with the
then existing experimental data (within a factor of two for more
than 90$\%$ (73$\%$) of nuclei with experimental half-lives shorter
than 1 s for $\beta^{-}$ ($\beta^{+}$/EC) decays). Most nuclei of
interest of astrophysical importance are the ones far from stability
and one has to rely on theoretical models to estimate their beta
decay properties. The accuracy of the pn-QRPA model increases with
increasing distance from the $\beta$-stability line (shorter
half-lives) \cite{Sta90,Hir93}. This is a promising feature with
respect to the prediction of experimentally unknown half-lives
(specially those present in the stellar interior), implying that the
predictions are made on the basis of a realistic physical model.
Later Nabi and Klapdor-Kleingrothaus  reported the calculation of
weak-interaction rates for more than 700 nuclei with A = 18 to 100
in stellar environment using the same nuclear model \cite{Nab99}.
These included capture rates, decay rates, gamma heating rates,
neutrino energy loss rates, probabilities of beta-delayed particle
emissions and energy rate of these particle emissions. The authors
then presented a detailed calculation of stellar weak-interaction
rates over a wide range of temperature and density scale for sd-
\cite{Nab99a} and fp/fpg-shell nuclei \cite{Nab04}. Since then these
calculations were further refined with use of more efficient
algorithms, computing power, incorporation of latest data from mass
compilations and experimental values, and fine-tuning of model
parameters both in the sd- shell (e.g. \cite{Nab08}) and fp-shell
nuclei (e.g. \cite{Nab05}).

The next section discusses the calculation of stellar
weak-interaction mediated rates using the pn-QRPA formalism. The
calculated Gamow-Teller strength distributions of waiting point
nuclei $^{72,74}$Kr and $^{76,78}$Sr are also presented in this
section. The rp-process electron capture and positron decay rates
are presented and compared with previous calculations in Section 3.
Summary and conclusions finally follow in Section 4.

\section {Description of the microscopic model}
The microscopic calculation of Gamow-Teller strength distributions
under stellar conditions is a challenging task. The calculations
become more challenging as the number of nucleons increases. Out of
the very limited available options, the pn-QRPA model can handle any
arbitrarily heavy system of nucleons and performs a microscopic
calculation of ground and excited state Gamow-Teller strength
functions \cite{Nab10,Nab10a}.

For the present pn-QRPA calculation single particle energies and
wave functions were calculated in the Nilsson model \cite{Nil55},
which takes into account nuclear deformations. Pairing was treated
in the BCS approximation. The proton-neutron residual interactions
occur in two different forms, namely as particle-hole and
particle-particle interactions. These interactions were given
separable form and were characterized by two interaction constants
$\chi$ and $\kappa$, respectively. These parameters are regarded as
the two most important model parameters in the pn-QRPA theory (see
Refs. \cite{Sta90,Hir93}). In this work a fine tuning of these
parameters was done which optimally reproduced the measured
half-lives for $^{72,74}$Kr and  $^{76,78}$Sr (for measured
half-lives see Refs. \cite{Piq03,Sch75,Gra92}). The
particle-hole-interaction parameter, $\chi$, is known to influence
the calculated position of the Gamow-Teller giant resonance (GTGR).
Normally for the case of stable nuclei, the constant $\chi$ is
parameterized such that the experimentally observed energy of the
GTGR is reproduced in the best way. However Hirsch and collaborators
\cite{Hir93}, using the pn-QRPA model,  deduced the locally best
$\chi$ values for individual isotopic chains. It was shown in
\cite{Hir93} that larger values of $\chi$ always shift GTGR to
higher excitation energy but the shift caused by a small change of
$\chi$ does not result in the same energetic shift for all isotopes.
On the other hand, the particle-particle-interaction ($\kappa$) is
known to enhance the ground-state correlations and leads to a
redistribution of the calculated $\beta$ strength (and is therefore
related to the half-lives of the nuclei), which is commonly shifted
to lower excitation energies \cite{Hir93}. Again, rather than going
for a parameterization of $\kappa$, the local best value for an
isotopic chain that best reproduces the measured half-lives, was
chosen in this work.  Fixing the interaction constants, $\chi$ and
$\kappa$, in this way led to considerable improvement in the
accuracy of the pn-QRPA calculation \cite{Sta90, Hir93, Nab99a,
Nab04}. For details of choosing the locally best values of the
strength parameters see Ref. \cite{Hir93}. The chosen values of
$\chi$ were 0.27 MeV and 0.36 MeV for Kr and Sr isotopes,
respectively. The corresponding $\kappa$ values were 0.045 Mev and
0.044 MeV, respectively.

The deformation parameter was recently argued as an important
parameter for QRPA calculations at par with the pairing parameter by
Stetcu and Johnson \cite{Ste04}. To make the problem more
challenging the nuclei in the mass region considered for this work
are known to have coexistence of prolate and oblate shapes. For
$^{72}$Kr the deformation parameter was calculated using the mass
formula of Ref. \cite{Moe81}. The chosen value of -0.2647 is also in
line with the findings of authors in Ref. \cite{Piq03,Pat93} which
favor an oblate shape for $^{72}$Kr. For the case of $^{74}$Kr and
$^{78}$Sr, the experimentally adopted values of the deformation
parameters, extracted by relating the measured energy of the first
$2^{+}$ excited state with the quadrupole deformation, were
available and taken from Raman et al. \cite{Ram87}. The deformation
parameter for the case of $^{76}$Sr was again taken from the mass
formula of Ref. \cite{Moe81} in close agreement with the findings of
N\'{a}cher and collaborators \cite{Nac04}. Table \ref{ta1} shows the
chosen values of nuclear deformations used in the model and the
calculated half-lives for the waiting point nuclei. All half-lives
are given in units of $sec$. Measured values are given in third
column whereas the pn-QRPA calculated terrestrial half-lives of Kr
and Sr isotopes are given in fourth column. The calculated
terrestrial half-lives have dominant contribution from positron
($\beta^{+}$) decay rates. Since the chosen values of $\chi$ and
$\kappa$ reproduced the measured half-lives fairly well, no
quenching factor was introduced in the present calculation. The last
column compares the pn-QRPA calculated stellar half-lives with the
terrestrial ones under limiting conditions. At the lowest
temperature considered in this work (T = 10$^{7}$ K), excited parent
states are not appreciably populated, while at the lowest density
considered in this work ($\rho Y_{e} = 10^{0.5}$ gcm$^{-3}$), the
continuum electron density is quite low. It is to be noted that
$\rho$ stands for the baryon density whereas $Y_{e}$ is the ratio of
the lepton number to the baryon number and their product $\rho
Y_{e}$ implies the electron density and is simply referred to as
density throughout this manuscript. Except for electron capture, the
stellar rates should be close to the terrestrial values. The
exception arises because only continuum electron capture and no
bound-states capture is calculated. Again, for the stated physical
conditions, the continuum electron capture rates are orders of
magnitude smaller than the corresponding positron ($\beta^{+}$)
decay rates. As can be seen the comparison is satisfactory. Nuclear
masses and Q-values required for the calculation were taken from the
mass compilation of Audi et al. \cite{Aud03}. The same set of model
parameters were used to calculate electron capture and positron
decay rates for these waiting point nuclei under rp-process
conditions.

The electron capture (ec) and positron decay (pd) rates of a
transition from the $i^{th}$ state of the parent to the $j^{th}$
state of the daughter nucleus are given by
\begin{eqnarray}\scriptsize
\lambda ^{^{ec(pd)} } _{ij} =\left[\frac{\ln 2}{D}
\right]\left[B(F)_{ij} +\left({\raise0.7ex\hbox{$ g_{A}
$}\!\mathord{\left/ {\vphantom {g_{A}  g_{V} }} \right.
\kern-\nulldelimiterspace}\!\lower0.7ex\hbox{$ g_{V}  $}}
\right)^{2} B(GT)_{ij} \right] \nonumber \\
\left[f_{ij}^{ec(pd)} (T,\rho Y_{e}  ,E_{f} )\right]. \label{phase
space}
\end{eqnarray}
The value of D was taken to be 6295s \cite{Yos88}. B(F) and B(GT)
are reduced transition probabilities of the Fermi and ~Gamow-Teller
(GT) transitions, respectively,
\begin{equation}
B(F)_{ij} = \frac{1}{2J_{i}+1} \mid<j \parallel \sum_{k}t_{\pm}^{k}
\parallel i> \mid ^{2}.
\end{equation}
\begin{equation}
B(GT)_{ij} = \frac{1}{2J_{i}+1} \mid <j \parallel
\sum_{k}t_{\pm}^{k}\vec{\sigma}^{k} \parallel i> \mid ^{2}.
\end{equation}
Here $\vec{\sigma}^{k}$ is the spin operator and $t_{\pm}^{k}$
stands for the isospin raising and lowering operator with
$(g_{A}/g_{V})$ = -1.254 \cite{Rod06}. Details of the calculation of
reduced transition probabilities (for both ground state and excited
states GT transitions) can be found in Ref. \cite{Nab99a}.

The $f_{ij}^{ec(pd)}$ are the phase space integrals and are
functions of stellar temperature ($T$), electron density ($\rho
Y_{e}$) and Fermi energy ($E_{f}$) of the electrons. They are
explicitly given by
\begin{equation}
f_{ij}^{ec} \, =\, \int _{w_{l} }^{\infty }w\sqrt{w^{2} -1}
 (w_{m} \, +\, w)^{2} F(+Z,w)G_{-} dw.
 \label{ec}
\end{equation}
and by
\begin{equation}
f_{ij}^{pd} \, =\, \int _{1 }^{w_{m}}w\sqrt{w^{2} -1} (w_{m} \,
 -\, w)^{2} F(- Z,w)(1- G_{+}) dw,
 \label{pd}
\end{equation}
In Eqs. ~(\ref{ec}) and ~(\ref{pd}), $w$ is the total energy of the
electron including its rest mass. $w_{m}$ is the total $\beta$-decay
energy,
\begin{equation}
w_{m} = m_{p}-m_{d}+E_{i}-E_{j},
\end{equation}
where $m_{p}$ and $E_{i}$ are masses and excitation energies of the
parent nucleus, and $m_{d}$ and $E_{j}$ of the daughter nucleus,
respectively. F($ \pm$ Z,w) are the Fermi functions and were
calculated according to the procedure adopted by Gove and Martin
\cite{Gov71}. G$_{\pm}$ are the Fermi-Dirac distribution functions
for positrons (electrons).
\begin{equation}
G_{+} =\left[\exp \left(\frac{E+2+E_{f} }{kT}\right)+1\right]^{-1},
\end{equation}
\begin{equation}
 G_{-} =\left[\exp \left(\frac{E-E_{f} }{kT}
 \right)+1\right]^{-1},
\end{equation}
here $E$ is the kinetic energy of the electrons and $k$ is the
Boltzmann constant.

The total capture/decay rate per unit time per nucleus is finally
given by
\begin{equation}
\lambda^{ec(pd)} =\sum _{ij}P_{i} \lambda _{ij}^{ec(pd)},
\label{total rate}
\end{equation}
where $P_{i}$ is the probability of occupation of parent excited
states and follows the normal Boltzmann distribution. After the
calculation of all partial rates for the transition $i \rightarrow
j$ the summation was carried out over 300 initial and final states
and satisfactory convergence was achieved in the rate calculation.
These 300 excited states were chosen with an appropriate bandwidth
covering energies up to 10 MeV in parent and 20 MeV in daughter
nuclei.

In order to ensure a reliable set of calculated rates three
important steps were undertaken in this calculation: (i) all
available experimental (XUNDL) data were incorporated in the
calculation; (ii) all available measured GT strength distribution
were inserted in the calculation wherever the theory was missing
them, and (iii) all excited state GT strength functions were
microscopically calculated within the pn-QRPA framework. The
theoretical excitation energy of a low-lying level was replaced by
the observed excitation energy if this experimental level had a
definite spin-parity assignment and was located within 0.5 MeV of
the theoretically calculated level. If more than one theoretically
calculated energy eigenvalue falls within 0.5 MeV of a measured
level then the replacement was done for the level lying closest to
the measured value. If there appeared a level without definite spin
and/or parity assignment in the sequence of experimental levels from
the ground state to higher excited levels, no replacement of
theoretical levels with measured levels was done beyond this
excitation energy. The calculated B(GT) value was then replaced by
the observed B(GT) value (if available) when the calculated energies
of the parent and daughter levels were replaced by the experimental
energies. Missing measured states were inserted and inverse
transitions (along with their log$ft$ values) were also incorporated
in the calculation. If the relevant transition was not measured but
the inverse transition was measured, the deduced observed B(GT)
value using this inverse transition was used in this calculation.
The measured B(GT$_{+}$) strength distribution of $^{72}$Kr from
Ref. \cite{Piq03}, of $^{74}$Kr from Ref. \cite{Poi04} and of
$^{78}$Sr from Ref. \cite{Nac04} were inserted in the calculation.
As mentioned earlier the pn-QRPA model has the additional advantage
that all excited state GT strength functions are calculated in a
microscopic fashion and the so-called Brink's hypothesis is not
assumed in this calculation. Brink's hypothesis states that GT
strength distribution on excited states is \textit{identical} to
that from ground state, shifted \textit{only} by the excitation
energy of the state. Recent pn-QRPA calculations (e.g.
\cite{Nab10,Nab10a}) have shown Brink's hypothesis to be a poor
approximation to be used in calculation of stellar weak rates
specially at high temperatures and densities. These steps ensured a
reliable set of calculation of ground and excited state GT strength
distributions to be later used for the calculation of weak rates for
waiting point nuclei.

\begin{figure}
\resizebox{0.75\textwidth}{!}{\includegraphics{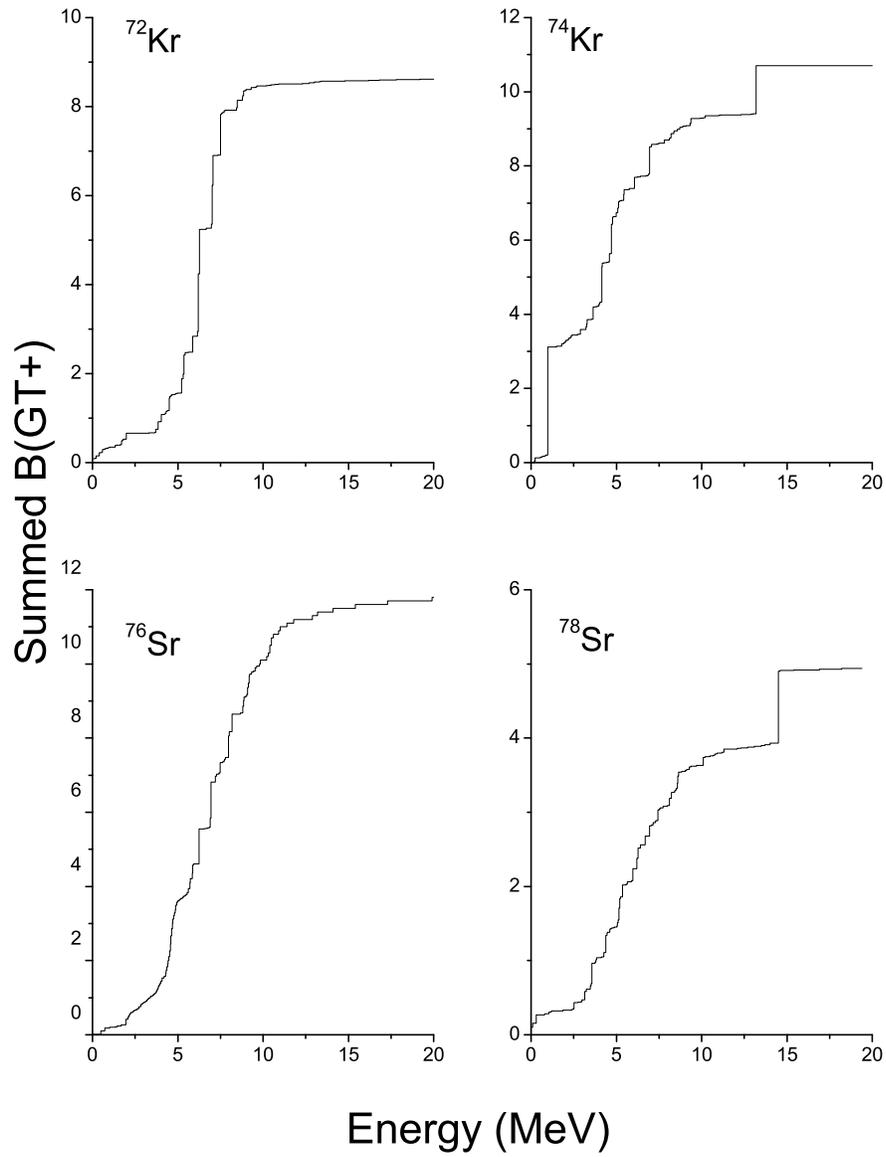}}
\caption{Cumulated Gamow-Teller strength distributions B(GT$_{+})$
for Kr and Sr isotopes. The abscissa represents energy in units of
MeV in daughter nucleus.} \label{fig1}
\end{figure}

Fig. \ref{fig1} shows the calculated BGT$_{+}$ distributions for the
Kr and Sr isotopes up to 20 MeV in daughter nuclei. As can be seen
from Fig. \ref{fig1}, the GT distributions are well fragmented and
extend up to high excitation energies in daughter nuclei. The model
independent Ikeda sum rule was satisfied in the calculation. The
discussion on the calculated weak rates follows in the next section.

\section{Thermal electron capture and positron decay rates of
waiting point nuclei}

\begin{figure}
\resizebox{0.75\textwidth}{!}{\includegraphics{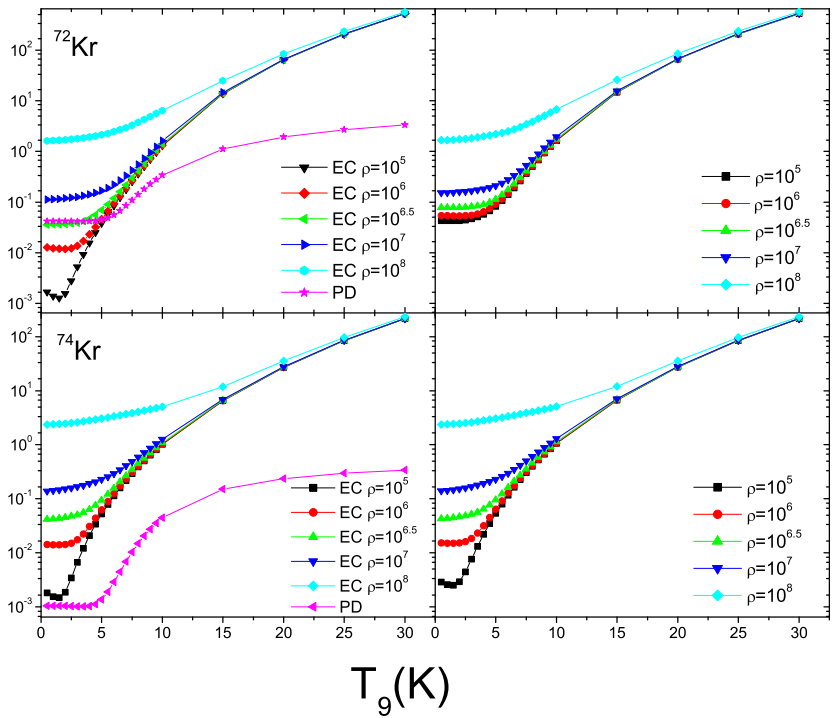}}
\caption{(Color online) Weak-interaction mediated rates for Kr
isotopes as a function of stellar temperature and density. The left
panels show the electron capture and positron decay rates whereas
the right panels show the combined rates. All rates are given in
units of $s^{-1}$. Densities are given in units of $gcm^{-3}$ and
temperatures in units of $10^{9}$ K.} \label{fig2}
\end{figure}

\begin{figure}
\resizebox{0.75\textwidth}{!}{\includegraphics{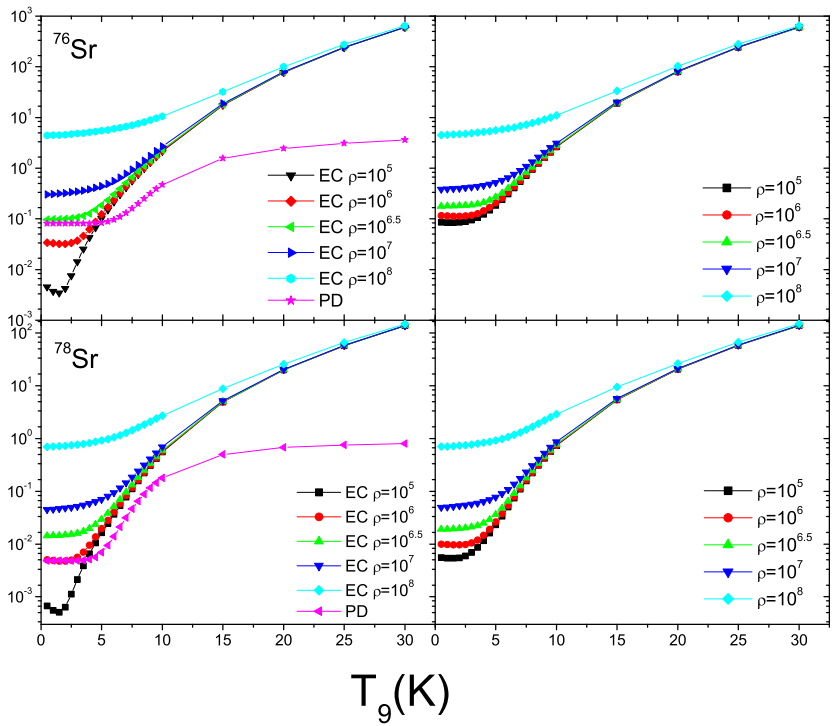}}
\caption{(Color online) Same as Fig.~\ref{fig2} but for Sr
isotopes.} \label{fig3}
\end{figure}

The calculated Fermi and GT strength distributions, along with the
corresponding phase space calculations were inserted into Eq.
~(\ref{phase space}) for the calculation of partial rates. The
calculated partial rates were finally added through the recipe of
Eq. ~(\ref{total rate}) to get the total stellar electron capture
and positron decay rates. The calculated rates for Kr and Sr
isotopes are shown in Fig. \ref{fig2} and Fig. \ref{fig3},
respectively.

The left panels of Fig. \ref{fig2} and Fig. \ref{fig3} depict the
calculated electron capture and positron decay rates, separately, as
a function of stellar temperature and density. The electron capture
rates are calculated at densities $\rho Y_{e} = 10^{5}, 10^{6},
10^{6.5}, 10^{7}$ and $ 10^{8}$ gcm$^{-3}$. T$_{9}$ gives the
stellar temperature in units of $10^{9}$ K. The calculation is
performed up to T$_{9}$ = 30. It is to be noted that such high
temperatures are far away from the relevant range of temperatures
for the rp-process. Nevertheless the calculation was performed till
T$_{9}$ = 30 in order to study the evolution of rates with
increasing stellar temperatures. The right columns in both figures
show the sum of calculated electron capture and positron decay rates
at the respective temperature and density points. The left panels
show that the calculated positron decay rates are almost independent
of stellar densities. It can be seen from the upper left panel of
Fig. \ref{fig2} that at low stellar temperatures (T$_{9} \leq 4 $K)
and densities ($\rho Y_{e} \leq 10^{6.5}$ gcm$^{-3}$), the positron
decay rates of $^{72}$Kr are bigger than the electron capture rates
by as much as a factor of 23. According to studies by authors in
Ref. \cite{Sch98,Wor94} the peak conditions for rp-process are in
the vicinity of T$_{9} = $ 1--3 K and $\rho Y_{e} = 10^{6}-10^{7}$
gcm$^{-3}$. It can be seen from the current calculation that
electron capture and positron decay rates of $^{72}$Kr contribute
roughly equally to the total weak rates at rp-process conditions.
However at high temperatures and densities the electron capture
rates shoot up and are bigger by more than two orders of magnitude
than the corresponding positron decay rates. For the case of
$^{74}$Kr the electron capture rates are always bigger than the
positron decay rates (lower left panel of Fig. \ref{fig2}). The
electron capture rates are more than 1--2 orders of magnitude bigger
than positron decay rates for rp-process conditions and the total
rates are commanded by the electron capture rates (lower right
panel). For still higher temperatures and densities the positron
decay rates are almost 3 orders of magnitude smaller and can safely
be neglected as compared to electron capture rates.

Fig. \ref{fig3} shows the calculated weak-interaction mediated rates
for $^{76,78}$Sr. The upper panels are for $^{76}$Sr and the lower
for $^{78}$Sr. Akin to the case of $^{72}$Kr, at low stellar
temperatures (T$_{9} \leq 4.5 $K) and densities ($\rho Y_{e} \leq
10^{6}$ gcm$^{-3}$), the positron decay rates dominate and are as
much as a factor 17 bigger than the corresponding electron capture
rates. At high temperatures and densities the electron capture rates
shoot exponentially due to increase in phase space factors and are
more than two orders of magnitude bigger. For the rp-process
conditions the situation is interesting. At density $\rho Y_{e} =
10^{6}$ gcm$^{-3}$ the positron decay rates are a factor 2--3
bigger, at $\rho Y_{e} = 10^{6.5}$ gcm$^{-3}$ the two rates are of
comparable magnitude while at $\rho Y_{e} = 10^{7}$ gcm$^{-3}$ the
electron capture rates are a factor 4 bigger. At high temperatures
and densities the total weak rates are driven by electron capture
rates (upper right panel of Fig. \ref{fig3}). For the case of
$^{78}$Sr the positron decay rates are bigger than the electron
capture rates by as much as a factor of 7 at low temperatures
(T$_{9} \leq 3.5 $K) and density ($\rho Y_{e} \leq 10^{5.5}$
gcm$^{-3}$) conditions. For the rp-process conditions the two rates
are of equal magnitude at $\rho Y_{e} = 10^{6}$ gcm$^{-3}$ whereas
the electron capture rates are bigger by a factor of 3--10 at $\rho
Y_{e} = 10^{6.5}-10^{7}$ gcm$^{-3}$. Once again at high temperatures
the electron capture rates are bigger by more than two orders of
magnitude as expected.

The contribution of parent excited states to the total electron
capture and positron decay rates is shown in Table \ref{ta2}. The
analysis was done at the selected stellar density of $\rho Y_{e} =
10^{6.5}$ gcm$^{-3}$ (a typical value for rp-conditions). Shown also
in Table \ref{ta2} are the ground state electron capture and
positron decay rates and the ratios of the ground state capture rate
to total electron capture rate ($R_{ec}(G/T)$) and ground state
decay rate to total positron decay rate ($R_{pd}(G/T)$) for the
waiting point nuclei. For the selected density scale no significant
contribution comes to the total decay rates from parent excited
states whereas at T$_{9} = 30 $K the excited states contribute
roughly 95$\%$ to the total capture rates. It is to be noted that
this analysis can change appreciably for higher stellar densities.

\begin{figure}
\resizebox{0.75\textwidth}{!}{\includegraphics{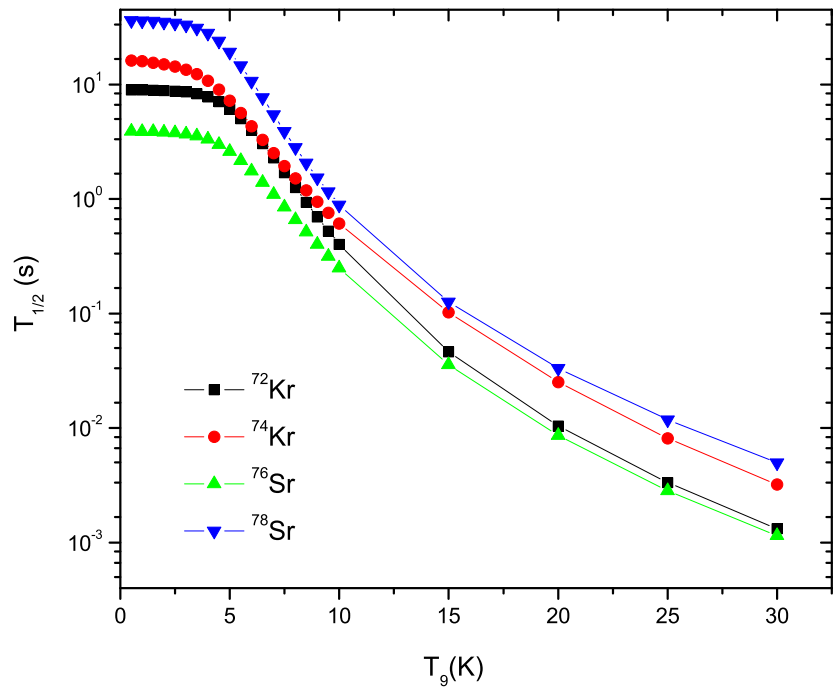}}
\caption{(Color online) Half-lives for Kr and Sr isotopes as a
function of stellar temperature calculated at a density of $ \rho
Y_{e} = 10^{6.5}$ gcm$^{-3}$.} \label{fig4}
\end{figure}

Fig. \ref{fig4} displays the half-lives (including both the positron
decay and electron capture contributions), as a function of stellar
temperatures at a selected density of $\rho Y_{e} = 10^{6.5}$
gcm$^{-3}$ for Sr and Kr waiting point nuclei. Again the calculation
is performed up to T$_{9} = 30 $K for reasons mentioned above. The
calculated half-lives decrease as stellar temperature increases as
expected since the electron capture rates increases substantially
with increasing temperatures (see Fig. \ref{fig2} and Fig.
\ref{fig3}). It is to be noted that for rp-process conditions the
half-lives decrease very mildly. For higher temperature the fall in
half-lives is rather steep.

The calculated rates were also compared with the  calculation of
weak-interaction rates for Kr and Sr isotopes by Sarriguren
\cite{Sar09}. The same author later expanded his calculation for
waiting point nuclei extending from Ni to Sn \cite{Sar11}.
Sarriguren used quasiparticle basis corresponding to a deformed
self-consistent Skyrme Hartee-Fock calculation with SLy4
\cite{Cha98} force. Pairing correlations were treated in BCS
formalism.  The spin-isospin interactions were considered both in
particle-hole and particle-particle channel and wavefunctions were
calculated using the Bohr-Mottelson factorization \cite{Boh75}.
Sarriguren used a quenching factor of 0.55 in his calculation. The
pn-QRPA calculated electron capture rates are in good agreement with
Sarriguren's rate except for the case of $^{74}$Kr. The pn-QRPA
positron decay rates are bigger specially at high temperatures by as
much as a factor of 40.

Table \ref{ta3} compares the weak-interaction rates and calculated
half-lives for Kr isotopes at $\rho Y_{e} = 10^{6}$ gcm$^{-3}$. In
the table $R_{ec}(N/S)$, $R_{pd}(N/S)$ and $R_{hl}(N/S)$ denote the
ratio of pn-QRPA calculated electron capture (ec), positron decay
(pd) and half-lives (hl), respectively, to those calculated by
Sarriguren. It can be seen from Table \ref{ta3} that the calculated
electron capture rates on $^{72}$Kr are in reasonable agreement with
those of Sarriguren. Positron decay rates of $^{72}$Kr are also in
reasonable agreement except at T$_{9} = 10 $K where reported rates
are around a factor 5 bigger. Sarriguren did not calculate
weak-interaction rates for T$_{9} > 10 $K. The last column shows the
ratio of the pn-QRPA calculated total  half-lives (including
electron capture and positron decay) to those calculated by
Sarriguren. The comparison of total half-lives for $^{72}$Kr is
correspondingly fairly good. Only at high temperatures is the
pn-QRPA calculated half life roughly half that reported by
Sarriguren. The origin of this difference might be connected to the
fact that energy levels beyond 1 MeV were not considered in
Sarriguren's calculation. These high-lying energy states have a
finite occupation probability at high temperatures and resulted in a
much bigger total weak rate. Much bigger differences are seen for
the case of weak rates for $^{74}$Kr. The reported electron capture
rates on $^{74}$Kr are a factor 3--7 bigger whereas positron decay
rates are almost double at low temperatures and around a factor 30
bigger at T$_{9} = 10 $K. The total weak rates are around an order
of magnitude bigger and consequently the calculated half-lives are
roughly an order of magnitude smaller. The reason for this
difference could be attributed to the different calculation of the
GT distributions in the two models. The pn-QRPA model calculated a
total B(GT) value of 10.73, up to 20 MeV, with the centroid lying
around 5.10 MeV in $^{74}$Br (see Fig. \ref{fig1}). The pn-QRPA
calculated total B(GT) strength value and/or placement of GT
centroid can lead to enhancements in reported rates.

A better comparison of reported calculation with that of Sarriguren
is seen for the case of Sr isotopes (Table \ref{ta4}). For $^{76}$Sr
the electron capture and positron decay rates are in reasonable
agreement. The calculated half-lives also agree fairly well except
at T$_{9} = 10 $K when the reported positron decay rate on $^{76}$Sr
is around a factor 4 bigger. The comparison of electron capture
rates on $^{78}$Sr is reasonable. The pn-QRPA reported positron
decay rates are double the Sarriguren's rate at low temperatures and
around factor 10 --40 bigger at higher temperatures. For the
rp-process conditions the pn-QRPA calculated electron capture rates
on $^{76}$Sr ($^{78}$Sr) are slightly bigger (smaller) and positron
decay rates for $^{76}$Sr ($^{78}$Sr) are slightly smaller (bigger)
than the corresponding Sarriguren rates. The calculated half-lives
of Sr isotopes are in reasonable agreement except at high
temperatures when the half-life calculated by Sarriguren is around a
factor 1.5 bigger for reasons already mentioned above.

\section{Conclusions}
To summarize weak-interaction rates for neutron-deficient Kr and Sr
isotopes were calculated within the framework of pn-QRPA model.
Proper choice of deformation parameter and smart choice of
particle-particle and particle-hole interaction terms led to correct
reproduction of the measured half-lives. The same model parameters
were used to calculate stellar weak rates  for rp-process
conditions.  Whereas electron capture and positron decay rates, of
other krypton and strontium isotopes in stellar matter, can be
calculated using the same model parameters as stated here, a case by
case fitting procedure  is required for other isotopic chains within
the framework of pn-QRPA model. Measured data were inserted wherever
available to increase the reliability of calculated rates. The
pn-QRPA reported electron capture rates on $^{74}$Kr are around a
factor 6 bigger than those calculated by Sarriguren under
rp-conditions. The present calculation clearly shows that the
electron capture rates are at least of similar magnitude as the
competing positron decay rates under rp-process conditions. The
calculation provide concrete evidence that for typical rp-process
conditions, T$_{9} = $ 1--3 K and $\rho Y_{e} = 10^{6}-10^{7}$
gcm$^{-3}$, electron capture rates on $^{72}$Kr and $^{76}$Sr are of
same order of magnitude as the corresponding positron decay rates.
On the other hand electron capture rates on $^{78}$Sr and $^{74}$Kr
are 1--2 orders of magnitude bigger than the respective positron
decay rates.  The study reconfirms the conclusion made by Sarriguren
that electron capture rates form an integral part of
weak-interaction mediated rates under rp-process conditions and
should not be neglected in nuclear reaction network calculations as
done in past (e.g. \cite{Sch98}).

\vspace{0.5 in}\textbf{Acknowledgments:} The author would like to
acknowledge the kind hospitality provided by the Abdus Salam ICTP,
Trieste, where this project was completed. The author wishes to
acknowledge the support of research grant provided by the Higher
Education Commission, Pakistan  through the HEC Project No. 20-1283.
The author also wishes to thank P. Sarriguren, A. Algora, E.
N\'{a}cher and B. Rubio for useful discussions and for providing
data used in this work.

\onecolumn
\begin{table}
\caption{Comparison of measured and pn-QRPA calculated half-lives in
units of $sec$. Nuclear deformations are given in second column. The
third and fourth columns give the measured and calculated
terrestrial half-lives, respectively. The fifth column gives the
limiting values of calculated stellar half-lives at density
$10^{0.5} g cm^{-3}$ and temperature $10^{7} K$.} \label{ta1}
\begin{center}
\begin{tabular}{ccccc} Nucleus & Deformation & $T_{1/2}$(exp) &
$T_{1/2}$(terr) & $T_{1/2}$(0.5,0.01)
\\\hline
$^{72}$Kr & -0.2647 & 17.2 & 17.2 & 16.7 \\
$^{74}$Kr & +0.3870 & 690 & 683.8 & 665.7 \\
$^{76}$Sr & +0.4074 & 8.9 & 8.8 & 8.53 \\
$^{78}$Sr & +0.4340 & 150 & 154.7 & 142.95 \\
\end{tabular}
\end{center}
\end{table}
\begin{table}
\caption{The \textit{ground state} electron and positron decay
rates, $\lambda_{ec}(G)$, $\lambda_{pd}(G)$, respectively, for
$^{72,74}Kr$ and $^{74,76}Sr$ in units of $sec^{-1}$. Given also are
the ratios of the ground state capture and decay rates to total
rate, $R_{ec}(G/T)$, $R_{pd}(G/T)$, respectively. The first column
gives the corresponding values of stellar density, $\rho Y_{e} $
($gcm^{-3}$), and temperature, $T_{9}$ (in units of $10^{9}$ K),
respectively.} \label{ta2}
\begin{center}
\begin{tabular}{c|cccc} & &\emph{$\mathbf{^{72}Kr}$} & & \\
$\mathbf{(\rho Y_{e} ,T_{9})}$& $\mathbf{\lambda_{ec}(G)}$ &
$\mathbf{R_{ec}(G/T)}$ &$\mathbf{\lambda_{pd}(G)}$ &
$\mathbf{R_{pd}(G/T)}$
\\\hline
($10^{6.5}$,0.01)& 3.57E-02 & 1.00E+00 &    4.15E-02 & 1.00E+00 \\
($10^{6.5}$,1)   & 3.59E-02 & 1.00E+00 &    4.15E-02 & 1.00E+00 \\
($10^{6.5}$,1.5) & 3.62E-02 & 1.00E+00 &    4.15E-02 & 1.00E+00 \\
($10^{6.5}$,2)   & 3.67E-02 & 1.00E+00 &    4.15E-02 & 1.00E+00 \\
($10^{6.5}$,2.5) & 3.75E-02 & 1.00E+00 &    4.15E-02 & 1.00E+00 \\
($10^{6.5}$,3)   & 3.91E-02 & 1.00E+00 &    4.15E-02 & 1.00E+00 \\
($10^{6.5}$,30)  & 1.48E+01 & 2.83E-02 &    2.24E-03 & 6.69E-04 \\
\hline & &\emph{$\mathbf{^{74}Kr}$} & & \\
($10^{6.5}$,0.01)& 4.15E-02 & 1.00E+00 &    1.04E-03 & 1.00E+00 \\
($10^{6.5}$,1)   & 4.25E-02 & 1.00E+00 &    1.04E-03 & 1.00E+00 \\
($10^{6.5}$,1.5) & 4.36E-02 & 1.00E+00 &    1.04E-03 & 1.00E+00 \\
($10^{6.5}$,2)   & 4.52E-02 & 1.00E+00 &    1.04E-03 & 1.00E+00 \\
($10^{6.5}$,2.5) & 4.74E-02 & 1.00E+00 &    1.04E-03 & 1.00E+00 \\
($10^{6.5}$,3)   & 5.05E-02 & 1.00E+00 &    1.04E-03 & 9.98E-01 \\
($10^{6.5}$,30)  & 1.16E+01 & 5.37E-02 &    3.26E-05 & 9.61E-05 \\
\hline & &\emph{$\mathbf{^{76}Sr}$} & & \\
($10^{6.5}$,0.01)& 9.61E-02 & 1.00E+00 &    8.12E-02 & 1.00E+00 \\
($10^{6.5}$,1)   & 9.69E-02 & 1.00E+00 &    8.12E-02 & 1.00E+00 \\
($10^{6.5}$,1.5) & 9.79E-02 & 1.00E+00 &    8.12E-02 & 1.00E+00 \\
($10^{6.5}$,2)   & 9.96E-02 & 1.00E+00 &    8.12E-02 & 1.00E+00 \\
($10^{6.5}$,2.5) & 1.02E-01 & 1.00E+00 &    8.12E-02 & 1.00E+00 \\
($10^{6.5}$,3)   & 1.07E-01 & 1.00E+00 &    8.12E-02 & 1.00E+00 \\
($10^{6.5}$,30)  & 2.36E+01 & 3.93E-02 &    4.42E-03 & 1.23E-03 \\
\hline & &\emph{$\mathbf{^{78}Sr}$} & & \\
($10^{6.5}$,0.01)& 1.42E-02 & 1.00E+00 &    4.85E-03 & 1.00E+00 \\
($10^{6.5}$,1)   & 1.43E-02 & 1.00E+00 &    4.85E-03 & 1.00E+00 \\
($10^{6.5}$,1.5) & 1.45E-02 & 1.00E+00 &    4.85E-03 & 1.00E+00 \\
($10^{6.5}$,2)   & 1.48E-02 & 1.00E+00 &    4.85E-03 & 1.00E+00 \\
($10^{6.5}$,2.5) & 1.53E-02 & 1.00E+00 &    4.85E-03 & 1.00E+00 \\
($10^{6.5}$,3)   & 1.60E-02 & 1.00E+00 &    4.84E-03 & 9.98E-01 \\
($10^{6.5}$,30)  & 3.59E+00 & 2.60E-02 &    1.41E-04 & 1.77E-04 \\
\end{tabular}
\end{center}
\end{table}
\begin{table}
\caption{Comparison of electron capture (ec) and positron decay (pd)
rates with those calculated by Sarriguren \cite{Sar09} for Kr
isotopes. Number in fourth and seventh column gives the ratio of
reported rates to the rates calculated by Sarriguren. The last
column shows the ratio of the pn-QRPA calculated total half-lives to
those calculated by Sarriguren. The rates are compared at stellar
density $\rho Y_{e}= 10^{6}$ gcm$^{-3}$. } \label{ta3}
\begin{center}\scriptsize
\begin{tabular}{c|ccccccc|c}  & & & & \emph{$\mathbf{^{72}Kr}$} & & & &\\
& & \textbf{ec} & & & & \textbf{pd} & &\\
$T_{9}$ &  Nabi & Sarriguren & $R_{ec}(N/S)$ & & Nabi & Sarriguren &
$R_{pd}(N/S)$ & $R_{hl}(N/S)$
\\\hline
0.01  & 1.29E-02 & 1.62E-02 & 7.97E-01 &  & 4.15E-02 & 3.47E-02 &  1.20E+00 &  9.35E-01\\
   1  & 1.23E-02 & 1.54E-02 & 7.99E-01 &  & 4.15E-02 & 3.48E-02 &  1.19E+00 &  9.33E-01\\
   2  & 1.19E-02 & 1.55E-02 & 7.66E-01 &  & 4.15E-02 & 3.99E-02 &  1.04E+00 &  1.04E+00\\
   3  & 1.37E-02 & 1.92E-02 & 7.12E-01 &  & 4.15E-02 & 5.07E-02 &  8.19E-01 &  1.27E+00\\
  10  & 1.33E+00 & 9.61E-01 & 1.38E+00 &  & 3.38E-01 & 7.52E-02 &  4.49E+00 &  6.21E-01\\ \hline
& & & & \emph{$\mathbf{^{74}Kr}$} & & & &\\
0.01  & 1.43E-02 & 2.20E-03 & 6.49E+00 &  & 1.04E-03 & 5.48E-04 &  1.90E+00 &  1.79E-01\\
   1  & 1.40E-02 & 2.20E-03 & 6.36E+00 &  & 1.04E-03 & 5.79E-04 &  1.80E+00 &  1.85E-01\\
   2  & 1.43E-02 & 2.66E-03 & 5.38E+00 &  & 1.04E-03 & 8.78E-04 &  1.18E+00 &  2.30E-01\\
   3  & 1.74E-02 & 3.65E-03 & 4.76E+00 &  & 1.03E-03 & 1.12E-03 &  9.20E-01 &  2.59E-01\\
  10  & 1.04E+00 & 3.32E-01 & 3.13E+00 &  & 4.44E-02 & 1.44E-03 &  3.07E+01 &
  3.08E-01\\
\end{tabular}
\end{center}
\end{table}
\begin{table}
\caption{Same as Table \ref{ta3} but for Sr isotopes.} \label{ta4}
\begin{center}\scriptsize
\begin{tabular}{c|ccccccc|c} & & & & \emph{$\mathbf{^{76}Sr}$} & & & &\\
& & \textbf{ec} & & & & \textbf{pd} & & \\
$T_{9}$ &  Nabi & Sarriguren & $R_{ec}(N/S)$ & & Nabi & Sarriguren &
$R_{pd}(N/S)$ & $R_{hl}(N/S)$
\\\hline
0.01  & 3.45E-02 & 3.04E-02 & 1.13E+00 &  & 8.12E-02 & 9.28E-02 &  8.75E-01 & 1.06E+00\\
   1  & 3.31E-02 & 2.99E-02 & 1.11E+00 &  & 8.12E-02 & 9.95E-02 &  8.16E-01 & 1.13E+00\\
   2  & 3.22E-02 & 3.04E-02 & 1.06E+00 &  & 8.12E-02 & 1.14E-01 &  7.10E-01 & 1.28E+00\\
   3  & 3.72E-02 & 3.55E-02 & 1.05E+00 &  & 8.11E-02 & 1.22E-01 &  6.63E-01 & 1.33E+00\\
  10  & 2.20E+00 & 1.55E+00 & 1.42E+00 &  & 4.73E-01 & 1.37E-01 &  3.46E+00 & 6.33E-01\\ \hline
& & & & \emph{$\mathbf{^{78}Sr}$} & & & &\\
0.01  & 5.09E-03 & 5.63E-03 & 9.04E-01 &  & 4.85E-03 & 2.46E-03 &  1.97E+00 & 8.14E-01\\
   1  & 4.88E-03 & 5.66E-03 & 8.63E-01 &  & 4.85E-03 & 2.75E-03 &  1.76E+00 & 8.64E-01\\
   2  & 4.78E-03 & 6.07E-03 & 7.87E-01 &  & 4.85E-03 & 3.35E-03 &  1.45E+00 & 9.78E-01\\
   3  & 5.59E-03 & 7.48E-03 & 7.47E-01 &  & 4.85E-03 & 3.59E-03 &  1.35E+00 & 1.06E+00\\
  10  & 5.72E-01 & 5.40E-01 & 1.06E+00 &  & 1.79E-01 & 4.60E-03 &  3.90E+01 & 7.25E-01\\
\end{tabular}
\end{center}
\end{table}



\end{document}